\documentclass{article}
\usepackage{amssymb}

\input{tcilatex}

\begin{document}

\title{{\large A droplet near the critical point: the divergence of Tolman's
length}\\
{\normalsize \ }}
\author{{\large M. A. Anisimov} \\
{\small Department of Chemical \& Biomolecular Engineering }\\
{\small \ \ and Institute for Physical Science \& Technology, }\\
{\small \ \ University of Maryland, College Park, MD 20742}}
\date{{\small 
\today%
}}
\maketitle

\begin{abstract}
{\small Application of "complete scaling" [Kim \textit{et al.}, Phys. Rev. E
67, 061506 (2003); Anisimov and Wang, Phys. Rev. Lett. 97, 25703 (2006)] to
the interfacial behavior of fluids shows that Tolman's length, a curvature
correction to the surface tension, diverges at the critical point of fluids
much more strongly than is commonly believed. The amplitude of the
divergence depends on the degree of asymmetry in fluid phase coexistence. A
new universal amplitude ratio, which involves this asymmetry, is
introduced.\ In highly asymmetric fluids and fluid mixtures the Tolman
length may become large enough near criticality to be detected in precise
experiments with microcapillaries and in simulations.}

{\small anisimov@umd.edu}
\end{abstract}

For decades the behavior of Tolman's length has remained one of the most
controversial issues in mesoscopic thermodynamics. The Tolman length ($%
\delta $) is defined as a curvature-correction coefficient in the surface
tension ($\sigma $) of a liquid or vapor droplet \cite{Tolman}:\newline
\begin{equation}
\sigma (R)=\sigma _{\infty }\left( 1-\frac{2\delta }{R}+...\right) ,
\end{equation}%
\newline
where $R$ is the droplet radius, taken equal to the radius of the surface of
tension \cite{RowlWid,Rowl94}, and $\sigma _{\infty }$ is the surface
tension for the planar interface. The physical origin of the curvature
correction is a difference between the equimolar surface and the surface of
tension; the difference is phenomenologically associated with asymmetry in
fluid phase coexistence. In symmetric systems, such as the lattice-gas model
and the regular-solution model, the difference between the equimolar surface
and the surface of tension vanishes and the Tolman length does not exist 
\cite{Rowl94,FisherWortis}. The curvature dependence of the surface tension
is important for the description of nucleation phenomena and fluids in
microcapillaries or nanopores. However, while square-gradient theories gave
consistent results in mean-field approximation \cite%
{FisherWortis,GiessenBukman,BlokhuisKuipers} , the actual critical behavior
of Tolman's length is not certain, either in sign or in behavior \cite%
{Phillips,BlokhuisKuipers}. Fisher and Wortis \cite{FisherWortis} and
Rowlinson \cite{Rowl84} have predicted a very weak algebraic divergence at
the critical point with an exponent $-0.065$. Since this prediction is
supported by an exact expression \cite{Rowl84} obtained for the
Widom-Rowlinson "penetrable-sphere model" \cite{WidRowl70}, it has become
commonly accepted.

In this Letter, I show that the application of \textquotedblleft complete
scaling\textquotedblright\ \cite{KimOrkoulas,AnisimovWang}, which properly
describes vapor-liquid asymmetry, to the problem of Tolman's length yields a
much stronger algebraic divergence at the critical point, with an exponent $%
-0.304$. The amplitude of this divergence depends on the degree of asymmetry
in fluid phase behavior. In highly asymmetric fluids and fluid mixtures this
divergence may be strong enough to be detected in accurate experiments and
simulations.

The generalized Laplace-Tolman equation reads \cite{RowlWid,Rowl94}.

\begin{equation}
P_{\alpha }-P_{\beta }=\frac{2\sigma _{\infty }}{R}\left( 1-\frac{2\delta }{R%
}+...\right) \text{,}
\end{equation}%
where $P_{\alpha }-P_{\beta }$ is the difference between the pressure inside
the droplet (phase $\alpha $) and outside (phase $\beta $). The difference $%
P_{\alpha }-P_{\beta }$ in Eq. (2) can be separated into two parts, namely,
a symmetric part, which does not depend on which fluid phase is inside or
outside the droplet, and an asymmetric part, \textit{i.e.}, $P_{\alpha
}-P_{\beta }=\left( P_{\alpha }-P_{\beta }\right) _{\text{\textrm{sym}}%
}+\left( P_{\alpha }-P_{\beta }\right) _{\text{\textrm{asym}}}$. Since
Tolman's length is associated with the asymmetric part only, one can
conclude that $2\sigma (\infty )/R=\left( P_{\alpha }-P_{\beta }\right) _{%
\text{\textrm{sym}}}$, and thus in first approximation obtain the expression%
\begin{equation}
\frac{2\delta }{R}\simeq -\frac{\left( P_{\alpha }-P_{\beta }\right) _{\text{%
\textrm{asym}}}}{\left( P_{\alpha }-P_{\beta }\right) _{\text{\textrm{sym}}}}%
\text{.}
\end{equation}

In scaling theory \cite{RowlWid} the surface tension in a three-dimensional
system can be estimated as a product of the "critical part" $\Delta P_{%
\mathrm{cr}}$ of the grand potential per unit volume ($\Omega /V=-P$), which
scales as $\Delta P_{\mathrm{cr}}\sim \xi ^{-3}$, and the characteristic
"thickness" of the interface, the correlation length of density fluctuations 
$\xi $. The correlation length diverges asymptotically as $\xi \approx \xi
_{0}^{-}\left\vert \Delta T\right\vert ^{-\nu }$ with the universal critical
exponent $\nu \simeq 0.630$ and the amplitude (below the critical point) $%
\xi _{0}^{-}$, while the surface tension vanishes at the critical point as $%
\sigma _{\infty }\approx \sigma _{0}\left\vert \Delta \hat{T}\right\vert
^{2\nu }\varpropto \xi ^{-2}$ \cite{scaling}. Here and below, the circumflex
indicates dimensionless variables, such as $\Delta \hat{T}=\left( T-T_{%
\mathrm{c}}\right) /T_{\mathrm{c}}$, $\Delta \hat{P}=\left( P-P_{\mathrm{c}%
}\right) /\rho _{\text{c}}k_{\text{\textrm{B}}}T_{\mathrm{c}}$, $\Delta \hat{%
\mu}=\left( \mu -\mu _{\text{\textrm{c}}}\right) /k_{\text{\textrm{B}}}T_{%
\text{\textrm{c}}}$, and $\Delta \hat{\rho}=\left( \rho -\rho _{\text{%
\textrm{c}}}\right) /\rho _{\text{c}}$, with $k_{\text{\textrm{B}}}$ being
Boltzmann's constant, while the subscript \textrm{c} indicates the critical
value of temperature, pressure, chemical potential, and molecular density,
correspondingly.

The function $\Delta P_{\mathrm{cr}}$ can also be separated into symmetric
and asymmetric parts. For a planar interface, $\sigma _{\infty }/\left(
\Delta P_{\mathrm{cr}}\right) _{\mathrm{sym}}\xi \approx c_{\sigma }$, where 
$c_{\sigma }\simeq 3.79$ \ is a universal constant. The mean-field value of $%
c_{\sigma }$ is $16/3$ \cite{anisimov}. Therefore, $\left( P_{\alpha
}-P_{\beta }\right) _{\text{\textrm{sym}}}R$ scales as $\left( \Delta P_{%
\mathrm{cr}}\right) _{\mathrm{sym}}\xi $ and $\left( P_{\alpha }-P_{\beta
}\right) _{\text{\textrm{asym}}}R^{2}$ scales as $\left( \Delta P_{\mathrm{cr%
}}\right) _{\mathrm{asym}}\xi ^{2}$. Hence, it follows from Eq. (4) that

\begin{equation}
\frac{2\delta }{\xi }\approx -c_{\delta }\frac{\left( \Delta P_{\mathrm{cr}%
}\right) _{\mathrm{asym}}}{\left( \Delta P_{\mathrm{cr}}\right) _{\mathrm{sym%
}}}\text{,}
\end{equation}%
where $c_{\delta }$ is another universal constant. As shown below, the
mean-field value of $c_{\delta }$ is $5/6$, while the scaling-theory value
is estimated as $\sim 0.6-0.7$.

Close to the critical point, $\Delta P_{\mathrm{cr}}$ can be approximated as 
$\Delta P_{\mathrm{cr}}\approx \chi \left( \Delta \mu \right) ^{2}$, where
the derivative $\chi =\left( \partial \rho /\partial \mu \right) _{T}=\left(
\partial ^{2}P/\partial \mu ^{2}\right) _{T}$, the susceptibility with
respect to density fluctuations, is taken for liquid or vapor phase
(depending on which would be considered as $\alpha $-phase) at the
corresponding phase boundary. The susceptibility $\chi $ becomes
asymptotically symmetric close to the critical point but exhibits growing
asymmetry upon departure from the critical point, containing two terms,
symmetric $\chi _{\text{\textrm{sym}}}$ (equal for both phases) and
asymmetric $\chi _{\text{\textrm{asym}}}$ (having opposite signs for the
different phases). Since $\left( \Delta P_{\mathrm{cr}}\right) _{\text{%
\textrm{sym}}}\approx \chi _{\text{\textrm{sym} }}\left( \Delta \mu \right)
^{2}$ and $\left( \Delta P_{\mathrm{cr}}\right) _{\text{\textrm{asym}}%
}\approx \chi _{\text{\textrm{asym}}}\left( \Delta \mu \right) ^{2}$, Eq.
(4) becomes%
\begin{equation}
\frac{2\delta }{\xi }\approx -c_{\delta }\frac{\chi _{\text{\textrm{asym}}}}{%
\chi _{\text{\textrm{sym}}}}\text{.}
\end{equation}

The vapor-liquid asymmetry in fluid criticality is most appropriately
treated by so-called "complete scaling", originally introduced by Fisher and
Orkoulas \cite{Fisher2000} and further elaborated by Kim \textit{et al.} 
\cite{KimOrkoulas} and by Anisimov and Wang \cite{AnisimovWang}.
Specifically, it was shown \cite{AnisimovWang} that with an appropriate
choice of the critical value of the entropy, $\rho _{\mathrm{c}}S_{\mathrm{c}%
}=\left( dP/dT\right) _{\text{\textrm{cxc}}}$, where the subscript \textrm{%
cxc} means vapor-liquid coexistence, the asymmetry in fluid criticality is
governed in first approximation by only two coefficients, $a_{3}$ and $%
b_{2}, $ in the linear mixing of three physical fields (pressure,
temperature, and chemical potential) into two independent scaling fields,
the "ordering" field $h_{1}$ and the \textquotedblleft
thermal\textquotedblright\ field $h_{2}$ according to

\begin{eqnarray}
h_{1} &=&\Delta \hat{\mu}+a_{3}\left[ \Delta \hat{P}-\left( \frac{d\hat{P}}{d%
\hat{T}}\right) _{\text{\textrm{cxc}}}\Delta \hat{T}\right] \text{,} \\
h_{2} &=&\Delta \hat{T}+b_{2}\Delta \hat{\mu}\text{,}
\end{eqnarray}%
where the derivative $\left( d\hat{P}/d\hat{T}\right) _{\text{\textrm{cxc}}}$
is taken at the critical point. The corresponding field-dependent
thermodynamic potential $h_{3}\left( h_{1},h_{2}\right) $ is also a linear
combination of the physical fields. In traditional (\textquotedblleft
incomplete\textquotedblright ) scaling for fluids, the nontrivial
coefficient $a_{3}$, responsible for the pressure mixing into the ordering
field, is assumed to be zero \cite{Mermin,AnisimovSeng}.

The total susceptibility $\chi =\chi _{\text{\textrm{sym}}}+\chi _{\text{%
\textrm{asym}}}$ is, in general, a combination of three symmetric scaling
susceptibilities: strong $\chi _{1}=\left( \partial ^{2}h_{3}/\partial
h_{1}^{2}\right) $, weak $\chi _{2}=\left( \partial ^{2}h_{3}/\partial
h_{2}^{2}\right) $, and cross $\chi _{12}=\left( \partial ^{2}h_{3}/\partial
h_{1}\partial h_{2}\right) $ \cite{KimOrkoulas,AnisimovSeng}. A detailed
derivation of the susceptibility, based on "complete-scaling", has been made
by Kim \textit{et al}. \cite{KimOrkoulas}. As a lower-order approximation of
the result given in ref. \cite{KimOrkoulas}, one can obtain%
\begin{eqnarray}
\hat{\chi} &=&\left( \frac{\partial \hat{\rho}}{\partial \hat{\mu}}\right) _{%
\hat{T}}=\hat{\chi}_{\text{\textrm{sym}}}+\hat{\chi}_{\text{\textrm{asym}}%
}\approx  \nonumber \\
&&(1+a_{3})^{2}\left( 1+\frac{3a_{3}}{1+a_{3}}\Delta \hat{\rho}\right) \chi
_{1}+b_{2}^{2}\chi _{2}+2(1+a_{3})b_{2}\chi _{12}\text{ .}
\end{eqnarray}%
When $a_{3}=0$ (traditional \textquotedblleft incomplete\textquotedblright\
scaling), Eq. (8) becomes identical to the result obtained in ref. \cite%
{AnisimovSeng}.

By using expressions for the scaling susceptibilities along the vapor-liquid
coexistence ($h_{1}=0$) obtained in refs. \cite{KimOrkoulas,AnisimovSeng},
one can derive from Eq. (8)

\begin{equation}
\hat{\chi}_{\text{\textrm{sym}}}\approx \Gamma _{0}^{-}\left\vert \Delta 
\hat{T}\right\vert ^{-\gamma }\left( 1+\Gamma _{1}^{-}\left\vert \Delta \hat{%
T}\right\vert ^{\theta }+...\right)
\end{equation}%
and%
\begin{equation}
\hat{\chi}_{\text{\textrm{asym}}}\approx \pm \Gamma _{0}^{-}\left\vert
\Delta \hat{T}\right\vert ^{-\gamma }\left[ \frac{3a_{3}}{1+a_{3}}%
B_{0}\left\vert \Delta \hat{T}\right\vert ^{\beta }-2b_{2}\frac{\beta B_{0}}{%
\Gamma _{0}^{-}}\left\vert \Delta \hat{T}\right\vert ^{1-\alpha -\beta }+...%
\right] \text{,}
\end{equation}%
where $\Gamma _{0}^{-}$ is the asymptotic critical amplitude of the
susceptibility, $B_{0}$ is the asymptotic amplitude in the expression $%
\Delta \hat{\rho}\approx \pm B_{0}\left\vert \Delta \hat{T}\right\vert
^{\beta }$, $\gamma \simeq 1.239$, $\beta \simeq 0.326$, and $\alpha \simeq
0.109$ are universal "Ising" critical exponents; $\Gamma _{1}^{-}$ is the
first symmetric ("Wegner") correction amplitude, and $\theta \simeq 0.5$ is
the "Wegner" correction exponent; $\pm $ corresponds to liquid and vapor
densities, respectively.

By expanding the ratio $2\delta /\xi \approx -c_{\delta }\left( \hat{\chi}_{%
\text{\textrm{asym}}}/\hat{\chi}_{\text{\textrm{sym}}}\right) $, one obtains%
\begin{equation}
\frac{2\delta }{\xi }\approx \mp c_{\delta }\left( \frac{3a_{3}}{1+a_{3}}%
B_{0}\left\vert \Delta \hat{T}\right\vert ^{\beta }-2b_{2}\frac{\beta B_{0}}{%
\Gamma _{0}^{-}}\left\vert \Delta \hat{T}\right\vert ^{1-\alpha -\beta
}+...\right) \left( 1-\Gamma _{1}^{-}\left\vert \Delta \hat{T}\right\vert
^{\theta }+...\right) \text{.}
\end{equation}%
Hence, Tolman's length contains two diverging terms, namely,%
\begin{equation}
\delta \approx \mp \xi _{0}^{-}c_{\delta }\left[ \frac{3a_{3}}{2\left(
1+a_{3}\right) }B_{0}\left\vert \Delta \hat{T}\right\vert ^{\beta -\nu
}-b_{2}\frac{\beta B_{0}}{\Gamma _{0}^{-}}\left\vert \Delta \hat{T}%
\right\vert ^{1-\alpha -\beta -\nu }+...\right] \text{,}
\end{equation}%
where $\mp $ corresponds to a liquid droplet or a bubble of vapor,
respectively.

The behavior of Tolman's length, given by Eq. (12), differs in an essential
way from the results obtained by previous investigators \cite%
{Rowl94,FisherWortis,Rowl84}: a new term, $\varpropto \left\vert \Delta \hat{%
T}\right\vert ^{\beta -\nu }$, emerges from the complete-scaling analysis.
The first term in Eq. (14) diverges more strongly, since $\beta -\nu \simeq
-0.304,$ whereas $1-\alpha -\beta -\nu \simeq -0.065$. In \textquotedblleft
incomplete scaling\textquotedblright , when $a_{3}=0$, Tolman's length
diverges very weakly, namely as$\left\vert \Delta \hat{T}\right\vert
^{1-\alpha -\beta -\nu }=\left\vert \Delta \hat{T}\right\vert ^{-0.065}$\cite%
{FisherWortis,Rowl84}. However, the \textquotedblleft incomplete scaling"
result may be valid for some specific systems, such as the \textquotedblleft
penetrable sphere model\textquotedblright\ \cite{Rowl84,WidRowl70}, since
that model has an exact symmetry axis on which the chemical potential is an
analytic function of temperature \cite{RenOrkoulas}.

Equation (12) can be further re-expressed as

\begin{equation}
\delta \approx \mp c_{\delta }\left[ \frac{3a_{3}}{2(1+a_{3})}\Delta \hat{%
\rho}+cb_{2}\frac{\Delta \left( \hat{\rho}\hat{S}\right) }{\Delta \hat{\rho}}%
\right] \xi \text{ ,}
\end{equation}%
with use of the universal ratio $c=\beta B_{0}^{2}/\Gamma
_{0}^{-}A_{0}^{-}\simeq 1.58$ \cite{Ising Ratios} and by introducing $\Delta
\left( \hat{\rho}\hat{S}\right) =\dint \left( C_{V}\right) _{\mathrm{cr}%
}dT/k_{\mathrm{B}}T\approx -\left( A_{0}^{-}/\left( 1-\alpha \right)
\left\vert \Delta \hat{T}\right\vert ^{1-\alpha }-B_{\mathrm{cr}}\left\vert
\Delta T\right\vert \right) $, the dimensional entropy (per unit volume)
deviation from its critical value; $A_{0}^{-}$ is the amplitude of the
critical (fluctuation-induced) part of the isochoric heat capacity $\left(
C_{V}\right) _{\mathrm{cr}}$ in the two-phase region \cite{AnisimovWang}.
The higher-order term $B_{\mathrm{cr}}\left\vert \Delta T\right\vert ,$
known as the "fluctuation-induced critical background", does not contribute
to the divergence of Tolman's length, since $1-\beta -\nu \simeq 0.044$.
However, this term is needed to satisfy crossover to the mean-field regime 
\cite{AnisimovWang}; moreover its contribution is of lower order than that
from the "Wegner" correction ($\beta +\theta -\nu \simeq 0.2$).

In mean-field approximation, where $c=2,$ $\beta =1/2,$ $\alpha =0,$ and $%
\nu =1/2$, Tolman's length remains finite:%
\begin{eqnarray}
\delta &\approx &\mp \bar{\xi}_{0}^{-}\overline{c}_{\delta }\left[ \frac{%
3a_{3}}{2\left( 1+a_{3}\right) }\bar{B}_{0}-2b_{2}\frac{\Delta \bar{C}_{V}}{%
k_{\text{\textrm{B}}}\bar{B}_{0}}\right]  \nonumber \\
&=&\mp \bar{\xi}_{0}^{-}\overline{c}_{\delta }\left[ \frac{3a_{3}}{2\left(
1+a_{3}\right) }\left( \frac{6\hat{\mu}_{11}}{\hat{\mu}_{30}}\right)
^{1/2}-2b_{2}\frac{3\hat{\mu}_{11}^{2}}{\hat{\mu}_{30}}\left( \frac{\hat{\mu}%
_{30}}{6\hat{\mu}_{11}}\right) ^{1/2}\right] \text{,}
\end{eqnarray}%
where $\bar{\xi}_{0}^{-}$ is the mean-field amplitude of the correlation
length and $\overline{c}_{\delta }$ is the mean-field value of the constant $%
c_{\delta }$; $\hat{\mu}_{ij}=\left( \partial ^{i+j}\hat{\mu}/\partial \hat{%
\rho}^{i}\partial \hat{T}^{j}\right) $ are the derivatives of the
dimensionless chemical potential, $\bar{B}_{0}=\left( \hat{\mu}_{30}/6\hat{%
\mu}_{11}\right) ^{1/2}$ is the mean-field amplitude of the density
coexistence curve, and $\Delta \bar{C}_{V}/k_{\text{\textrm{B}}}=3\hat{\mu}%
_{11}^{2}/\hat{\mu}_{30}$ is the mean-field heat-capacity discontinuity \cite%
{Croxton}. The leading asymmetry coefficients, which are assumed to be
unaffected by fluctuations, can also be expressed through the derivatives $%
\hat{\mu}_{ij}$\cite{AnisimovWang} as%
\begin{equation}
\frac{a_{3}}{1+a_{3}}=\frac{2}{3}\frac{\hat{\mu}_{21}}{\hat{\mu}_{11}}-\frac{%
1}{5}\frac{\hat{\mu}_{40}}{\hat{\mu}_{30}},\text{ \ \ \ \ \ \ \ }b_{2}=\frac{%
1}{\hat{\mu}_{11}}\left( \frac{\hat{\mu}_{21}}{\hat{\mu}_{11}}-\frac{1}{5}%
\frac{\hat{\mu}_{40}}{\hat{\mu}_{30}}\right) \text{.}
\end{equation}%
Substituting Eqs. (15) into Eq. (14), one obtains the compact result%
\begin{equation}
\frac{\delta }{\bar{\xi}_{0}^{-}}=\pm \frac{\overline{c}_{\delta }}{10}\bar{B%
}_{0}\frac{\hat{\mu}_{40}}{\hat{\mu}_{30}}=\pm \frac{\overline{c}_{\delta }}{%
10}\left( \frac{6\hat{\mu}_{11}}{\hat{\mu}_{30}}\right) ^{1/2}\frac{\hat{\mu}%
_{40}}{\hat{\mu}_{30}}\text{.}
\end{equation}%
By comparing this with the mean-field result of Fisher and Wortis \cite%
{FisherWortis} and Giessen \textit{et al.} \cite{GiessenBukman}, namely, $%
\delta /\bar{\xi}_{0}^{-}=\pm \bar{B}_{0}\hat{\mu}_{40}/12\hat{\mu}_{30}=\pm
\left( 6\hat{\mu}_{11}\hat{\mu}_{30}\right) ^{1/2}\hat{\mu}_{40}/12\hat{\mu}%
_{30}$, one obtains the mean-field value $\overline{c}_{\delta }=5/6$. For
the (mean-field) van der Waals fluid, where $a_{3}=1/4$, $b_{2}=4/45$, $\hat{%
\mu}_{21}=0$, $\hat{\mu}_{11}=9/4$, $\hat{\mu}_{40}=-\hat{\mu}_{30}=-27/8$,
and $\bar{B}_{0}=2$ \cite{AnisimovWang}, one obtains $\delta =\mp \bar{\xi}%
_{0}^{-}/6$. If one adopts $\overline{c}_{\delta }=1$, $\delta =\mp \bar{\xi}%
_{0}^{-}/5$. Hence, Tolman's length in the van der Waals fluid is negative
for a drop of liquid and positive for a bubble of vapor, in agreement with
the conclusions of Fisher and Wortis \cite{FisherWortis} and Blokhuis and
Kuipers \cite{BlokhuisKuipers}. The same sign is anticipated for real
asymmetric fluids when $a_{3}$ is large and positive \cite{AnisimovWang}.

It is also interesting that $\hat{\mu}_{21}$does not enter in the mean-field
result for Tolman's length, which depends only on one asymmetry amplitude $%
\hat{\mu}_{40}$, the fifth-order coefficient in the expansion of the
Helmholtz energy per unit volume in powers of density. When $a_{3}=0,$ Eq.
(15) gives $b_{2}=\hat{\mu}_{40}/10\hat{\mu}_{30}\hat{\mu}_{11}$ and the
mean-field result for Tolman's length given by Eq. (16) remains unchanged.
However, in the scaling regime this asymmetry term, renormalized by
fluctuations, produces two different singularities, namely, $\propto
\left\vert \Delta \hat{T}\right\vert ^{\beta -\nu }$ and $\propto \left\vert
\Delta \hat{T}\right\vert ^{1-\alpha -\beta -\nu }$. A renormalization-group
treatment of the fifth-order term in the initial Hamiltonian, with this
fifth-order term regarded as a \textquotedblleft non-Ising
asymmetry\textquotedblright , produces a new critical exponent $\theta
_{5}\cong 1.3$ \cite{Zhang} and results in a nondivergent contribution to
Tolman's length, $\propto \left\vert \Delta \hat{T}\right\vert ^{\theta
_{5}-\nu }\simeq \left\vert \Delta \hat{T}\right\vert ^{0.7}$ \cite%
{FisherWortis}. Actually, the fifth-order term in the initial Hamiltonian
also generates a singularity, $\propto \left\vert \Delta \hat{T}\right\vert
^{\beta -\nu }$, making a special treatment of \textquotedblleft non-Ising
asymmetry\textquotedblright\ practically irrelevant.

In order to obtain an estimate for the universal constant $c_{\delta }$ in
the scaling regime, \ one can \ consider an approximate
square-gradient-theory expression for Tolman's length \cite{BlokhuisKuipers}%
, namely,%
\begin{equation}
\delta \simeq -\frac{\sigma _{\infty }\left( \chi _{\alpha }-\chi _{\beta
}\right) }{\left( \rho _{\alpha }-\rho _{\beta }\right) ^{2}}\text{.}
\end{equation}%
By calculating the difference of the susceptibilities via Eqs. (9) and (10),
and by using $\left( \rho _{\alpha }-\rho _{\beta }\right) ^{2}\approx
4B_{0}\left\vert \Delta \hat{T}\right\vert ^{\beta }$, $\sigma _{\infty
}\approx \sigma _{0}\left\vert \Delta \hat{T}\right\vert ^{2\nu }$ and the
universal ratios $\sigma _{0}/k_{\mathrm{B}}T_{c}\rho _{\mathrm{c}}\xi
_{0}^{-}A_{0}^{-}\simeq 2.25$ \cite{anisimov}, $\Gamma
_{0}^{-}A_{0}^{-}/B_{0}^{2}\simeq 0.21$ \cite{Ising Ratios}, one obtains $%
c_{\delta }\simeq 0.47$. Since in mean-field approximation Eq. (17) gives $%
\overline{c}_{\delta }=2/3$ instead of the "exact" value $\overline{c}%
_{\delta }=5/6$ \cite{FisherWortis,BlokhuisKuipers} or instead of (also
plausible) $\overline{c}_{\delta }=1$, the $c_{\delta }$ value must be
corrected at least by a factor of $5/4$, yielding $c_{\delta }\simeq 0.6$ or
by$\ 3/2$, yielding $c_{\delta }\simeq 0.7$.

One can also notice that Tolman's length can be thermodynamically expressed
through the amount of excess adsorption, $\Gamma _{\mathrm{s}}$, at the
surface of tension as $\delta =\Gamma _{\mathrm{s}}/\left( \rho _{\alpha
}-\rho _{\beta }\right) $ \cite{Tolman,BlokhuisKuipers}. Assuming that \ the
excess (asymmetric part) density at the surface of tension is associated
with the "singular diameter" (the deviation of the mean density from the
critical isochore), $\hat{\rho}_{\mathrm{d}}-1\equiv \left( \rho _{\alpha
}+\rho _{\beta }\right) /2\rho _{\mathrm{c}}-1\approx a_{3}/(1+a_{3})\left(
\Delta \hat{\rho}\right) ^{2}+b_{2}\Delta \left( \hat{\rho}\hat{S}\right) $ 
\cite{AnisimovWang}, in such a way that $\Gamma _{\mathrm{s}}\approx -\xi
\rho _{\mathrm{c}}\left( \hat{\rho}_{\mathrm{d}}-1\right) $, one obtains $%
\delta \approx -\xi \left( \hat{\rho}_{\mathrm{d}}-1\right) /\left( \rho
_{\alpha }-\rho _{\beta }\right) $. If one adopts $c_{\delta }=2/3,$ this
estimate asymptotically agrees with Eq. (13). For the van der Waals fluid
such an assumption yields $\delta /\bar{\xi}_{0}^{-}\cong \mp D\overline{/B}%
_{0}=\mp 1/5$,\ where $D=\left( \hat{\rho}_{\mathrm{d}}-1\right) /\left\vert
\Delta \hat{T}\right\vert =-\left( 3/5\right) \hat{\mu}_{11}\hat{\mu}_{40}/%
\hat{\mu}_{30}^{2}=2/5$ is the mean-field rectilinear diameter of the
coexistence densities \cite{Croxton}, in agreement with Eq. (16) if $%
\overline{c}_{\delta }=1$.

The exponent combination $\beta -\nu $, which is responsible for the leading
divergence of the Tolman's length, arises in several contexts in surface
thermodynamics \cite{Widom}. In particular, the total adsorption $\Gamma
\sim \xi \left( \rho _{\alpha }-\rho _{\beta }\right) $ diverges at the
critical point as $\left\vert \Delta \hat{T}\right\vert ^{\beta -\nu }$.
Interestingly, in mean-field approximation, where, formally, $\beta -\nu =0$%
, the adsorption, instead of being finite, becomes logarithmically divergent 
\cite{RowlWid,Widom}. Whether a similar logarithmic divergence may also
persist in the mean-field approximation for Tolman's length in the critical
region remains unclear.

A straightforward application of the present results to the problem of
Tolman's length in asymmetric binary fluids can be made by a generalization
of "complete scaling" \cite{Cerdeirina} with use of the "isomorphism
principle" \cite{AnisimovSeng}. In particular, one can use the expression
for Tolman's length given by Eq. (13) for "incompressible" liquid mixtures
replacing $\Delta \hat{\rho}$ by $\Delta \hat{x}=\left( x-x_{\mathrm{c}%
}\right) /x_{\mathrm{c}},$ the deviation of the molar fraction $x$ from its
critical value $x_{\mathrm{c}}$.

The divergence of Tolman's length may be large enough to be detected in well
designed accurate experiments with microcapillaries and in simulations.
Predictions for Tolman's length in several fluids exhibiting strong
vapor-liquid asymmetry are shown in Fig. 1. In less asymmetric fluids with
small-size molecules, such as nitrogen or methane, the coefficient $a_{3}$
is too small \cite{AnisimovWang} to make Tolman's length growing to a
nanoscale in any reasonable proximity to the critical point. However, in
highly asymmetric fluids and fluid mixtures, such as high-molecular-weight
hydrocarbons and their mixtures with small-molecular-volume species, or in
ionic fluids, the Tolman's length \ can easily reach a nanoscale. From Eq.
(13) with $c_{\delta }=2/3$, for the liquid mixture \textit{n}%
-hexadecane+nitrobenzene, one obtains $-2\delta \simeq 5$ nm at $\left\vert
\Delta \hat{T}\right\vert =10^{-4}$, while $\xi \approx \xi
_{0}^{-}\left\vert \Delta T\right\vert ^{-0.63}\simeq 60$ nm. For the
so-called "restricted primitive model" (RPM) of a electrolyte, which
exhibits a vapor-liquid phase transition in a system of oppositely charged
spherical ions of the same size, the Tolman's length is predicted to be
twice as large. The correlation length must be sufficiently smaller than the
size of the drop, allowing a thermodynamic approach while retaining the
concept of the droplet interface. Therefore, in these systems, for a
micron-size drop, which at $\left\vert \Delta \hat{T}\right\vert =10^{-4}$
is still much larger than the correlation length/thickness of the interface,
the curvature correction to the surface tension is about 1-2\%. Faraway from
the critical point, Tolman's length is only a small fraction of the
molecular size and, in order to obtain a 1-2\% correction to the surface
tension, one needs a nanosize drop.

In conclusion, let me emphasize the principal result given by Eq. (13) and
presented in Fig. 1. Tolman's length diverges at the critical point much
more strongly than is commonly believed. In fluids exhibiting strong
asymmetry of phase coexistence, Tolman's length may become mesoscopic near
criticality, thus noticeably affecting the surface tension of curved
interfaces.

Valuable discussions with Edgar M. Blokhuis, Michael E. Fisher, Michael R.
Moldover, Jan V. Sengers, Jan Thoen, and Benjamin Widom are highly
appreciated. I am also thankful to Heather St.Pierre for assistance.

\bigskip

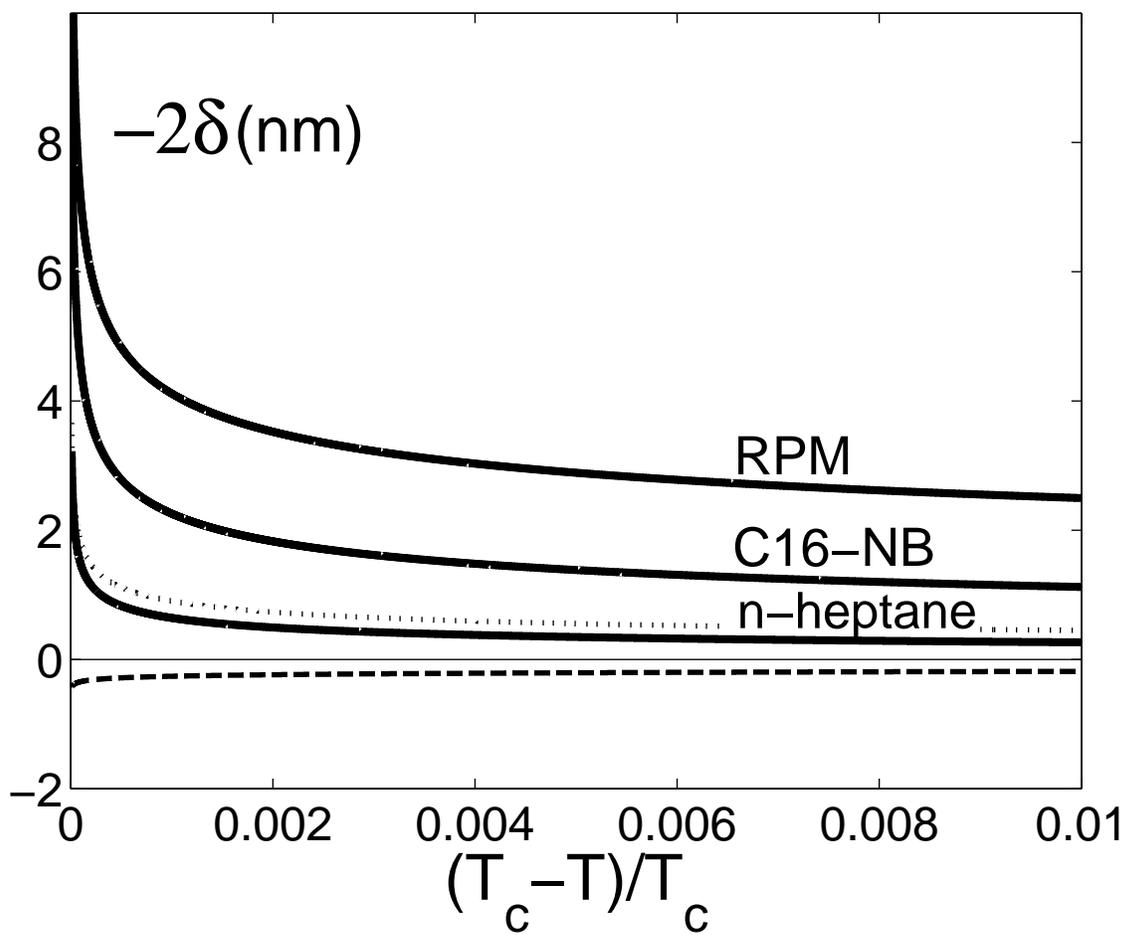
\textbf{Fig. 1.} Tolman's length (multiplied by factor of 2), calculated
from Eq. (13) with $c_{\delta }=2/3$, as a function of distance to the
critical temperature. The solid curves represent \textit{n}-heptane ($\xi
_{0}^{-}\simeq 0.11$ nm, $a_{3}\simeq 0.37$, $b_{2}\simeq 0.09$, $%
B_{0}\simeq 1.84$, $A_{0}^{-}\simeq 22.6$, and $B_{\mathrm{cr}}\simeq 18.4$ 
\cite{AnisimovWang,Wang}), \textit{n}-hexadecane+nitrobenzene (C16-NB) ($\xi
_{0}^{-}\simeq 0.18$ nm, $a_{3}\simeq 0.48$, $b_{2}\simeq 0$, and $%
B_{0}\simeq 2.4$, \cite{Cerdeirina,Wang}), and the restricted primitive
model (RPM) ($\xi _{0}^{-}\simeq 0.18$ nm for the $0.5$ nm ion diameter, $%
a_{3}\simeq 0.14$, $b_{2}\simeq -0.48$, $B_{0}\simeq 3.64$, $A_{0}^{-}\simeq
11.6$, and $B_{\mathrm{cr}}\simeq 4.9$ \cite{AnisimovWang,Wang}). The dotted
curve represents the positive contribution to $-2\delta $ for \textit{n}%
-heptane, diverging as $\left\vert \Delta \hat{T}\right\vert ^{\beta -\nu },$
while the dashed curve is the negative contribution, diverging as $%
\left\vert \Delta \hat{T}\right\vert ^{1-\alpha -\beta -\nu }.$


\begin{thebibliography}{10}
\bibitem[12]{scaling} In symmetric systems, along the phase coexistence, $%
\Delta \hat{P}_{\mathrm{cr}}=\left( \Delta \hat{P}_{\mathrm{cr}}\right) _{%
\mathrm{sym}}\approx A_{0}^{-}\left\vert \Delta \hat{T}\right\vert
^{2-\alpha }/\left( 2-\alpha \right) \left( 1-\alpha \right) \varpropto \xi
^{-3}$, where $2-\alpha =3\nu $ and $A_{0}^{-}$ is the asymptotic amplitude
of the isochoric heat capacity in the two-phase region.
\end{thebibliography}
\end{document}